\documentclass[conference]{IEEEtran}
\IEEEoverridecommandlockouts

\usepackage{cite}
\usepackage{amsmath,amssymb,amsfonts}
\usepackage{algorithmic}
\usepackage{graphicx}
\usepackage{textcomp}
\usepackage{xcolor}
\usepackage{minted}
\usepackage{booktabs}
\usepackage{tcolorbox}
\usepackage{array}
\usepackage{tabularx}
\usepackage{makecell}

\usepackage{amsthm}
\usepackage{xcolor} 
\usepackage{hyperref} 
\usepackage[normalem]{ulem} 

\def\BibTeX{{\rm B\kern-.05em{\sc i\kern-.025em b}\kern-.08em
    T\kern-.1667em\lower.7ex\hbox{E}\kern-.125emX}}

\newcommand{\SystemName}{xRoute}

\begin{document}

\title{Towards Policy-Enabled Multi-Hop Routing for Cross-Chain Message Delivery}

\author{
\IEEEauthorblockN{Amin Rezaei, Solomon L. Davidson, Bernard Wong}
\IEEEauthorblockA{
\textit{David R. Cheriton School of Computer Science} \\
\textit{University of Waterloo} \\
Waterloo, ON, Canada \\
\texttt{\{a27rezae,solomon.davidson,bernard\}@uwaterloo.ca}
}
}

\maketitle

\begin{abstract}

Blockchain ecosystems face a significant issue with liquidity fragmentation, as applications and assets are distributed across many public chains with each only accessible by subset of users. Cross-chain communication was designed to address this by allowing chains to interoperate, but existing solutions limit communication to directly connected chains or route traffic through hubs that create bottlenecks and centralization risks.

In this paper, we introduce \textbf{xRoute}, a cross-chain routing and message-delivery framework inspired by traditional networks. Our design brings routing, name resolution, and policy-based delivery to the blockchain setting. It allows applications to specify routing policies, enables destination chains to verify that selected routes satisfy security requirements, and uses a decentralized relayer network to compute routes and deliver messages without introducing a trusted hub. Experiments on the chains supporting the Inter-Blockchain Communication (IBC) protocol show that our approach improves connectivity, decentralization, and scalability compared to hub-based designs, particularly under heavy load.

\end{abstract}

\section{Introduction}

Blockchain adoption has grown at a staggering rate over the past decade. What started as a way of securely transferring assets between end users without relying on banks, exchanges, or other trusted intermediaries has since expanded into a rich ecosystem spanning thousands of independent chains, with many supporting sophisticated financial transactions. This shift to a multi-chain ecosystem was driven in part by conflicting application requirements that cannot be supported by a single chain, and it also alleviates congestion by allowing chains to transact independently.

With the emergence of a multi-chain ecosystem, providing interoperability across chains has become increasingly more important. The current most commonly used interoperability solution is to bridge assets, such as ERC-20 tokens, across chains by locking up assets on the source chain using a smart contract, and then minting a representation of the locked asset on the destination chain.
According to DeFiLlama~\cite{defillama}, as of \emph{Jan. 2026}, more than \$45~billion worth of tokens are locked on cross-chain bridges.

However, securing blockchain bridges has proven to be challenging as more than 2.8 billion dollars have been stolen from bridges~\cite{bridge-vulnerabilities}. In practice, most bridges rely on smart contracts controlled by a small committee of operators, requiring users to trust this group. Many of the bridge exploits used social engineering techniques to acquire the small number of keys needed to take control of the bridge's smart contract. Furthermore, most bridges only provide interoperability between two chains. Bridges between smaller chains often do not exist, or have questionable reputation. As a result, users of smaller chains often need to bridge their assets more than once across possibly low-reputation bridges in order for their assets to be available at their desired destination, which significantly increases their risk exposure.

The Inter-Blockchain Communication (IBC) protocol was designed to avoid some of the security issues that are present in smart contract-based bridges. Instead of relying on a smart contract that can be controlled by a small number of signatories, IBC enables two chains to communicate through light clients that track the block headers and verify the consensus states of counter-party chains. 
Subsequently, inclusion of transactions can be easily verified through Merkle proofs. This provides a comparatively secure approach since it offers equivalent security to that offered by the other chain, assuming proper bootstrapping of clients.

Unfortunately, out of the chains that support IBC, most only have IBC connections to a few popular chains. This limitation is driven in part by the high administrative and financial overhead to create and maintain IBC connections. While a multi-hop IBC extension~\cite{cosmos_ibc_multihop} exists that allows messages to traverse intermediate chains, it does not specify how routes should be computed or how security policies should be enforced. Users who wish to communicate between chains where an IBC connection does not exist must find their own multi-chain route and send a series of messages that must interact with wallets or smart contracts on the intermediate chains. Not only is this approach challenging to undertake, it can also be problematic from a security perspective as the user may unknowingly choose to route through a malicious chain that can compromise the message. Furthermore, with limited information about the gas cost of different chains, the user may end up choosing a high-fee route. 

Axelar~\cite{axelar-wp} is a cross-chain messaging solution that addresses this issue by relaying all cross-chain messages through their own chain to create a hub-and-spoke topology. Although this provides connectivity between all chains connected to the hub chain, it creates a new significant trust assumption. Furthermore, the hub chain can limit cross-chain messaging throughput, which can in turn increase message delivery cost as transaction fees increase with demand.

In many ways, cross-chain communication today is facing very similar problems to those that the Internet had faced and solved for inter-network communication. This includes the need for decentralized route computation that meets application-specific security requirements, and efficient delivery of messages without introducing new trust assumptions. At the same time, many of these problems have unique blockchain-specific assumptions and requirements that demand solutions that differ from existing networking solutions. Solving these cross-chain communication problems may enable blockchains to experience the same transformative growth by the general public that the Internet had experienced.

In this paper, we introduce \SystemName{}, a decentralized message routing and delivery framework built on top of multi-hop IBC. \SystemName{} constructs routes over existing direct connections without introducing additional trust assumptions. These routes meet user-specified policies; 
We distinguish between (1) security policies, which must hold and are enforced on-chain such as a minimum Nakamoto coefficient~\cite{srinivasan2017quantifying} for intermediate chains, where the coefficient is a measure of the minimum number of independent entities that can collude to shut down a blockchain, and (2) preference policies, which should hold for best user experience (e.g., fee minimization, latency minimization). \SystemName{} enforces security policies at the destination, while off-chain relayers (coordinated via a Relayer Network) optimize preference policies such as fees and latency and improve user experience.

Due to the cost of performing on-chain computation, \SystemName{} incentivizes third-party relayers (parties responsible for watching chains for outgoing IBC packets and delivering them) to perform route computation off-chain. However, routes provided by the relayers are verified on-chain by the destination to meet the security requirements before they are used. Additionally, the relayers are further incentivized to provide routes that optimize for non-security metrics, such as minimizing transaction fees or using chains with low finalization times.

In order for a single relayer to relay a message across a multi-hop IBC connection, it must be running light clients for every chain in the path, and it must also have gas tokens for the intermediate hops and the destination to pay for transaction fees. This can be a significant administrative barrier in deploying multi-hop relayers. Instead, \SystemName{} introduces a relayer network organized as a separate blockchain, where relayers specializing in specific chains cooperatively perform both route computation and message delivery. With this design, each relayer only needs to hold gas tokens for a single chain, and additional relayers can be introduced to the network to handle increased relayer load driven by user demand.

Our experimental results show (i) near 90\% connectivity vs. 15\% for hub-and-spoke, (ii) more than 30\% connectivity (50\% with topology upgrades) after removing the top four chains, (iii) less than \$0.10 on-chain cost per message and, (iv) low processing time even with more than $16k$ transactions/s.

Overall, this paper makes three contributions:

\begin{enumerate}
  \item \textbf{Policy-driven routing over multi-hop IBC.} We design \SystemName{}, a framework that strictly enforces security policies at the destination without introducing a new trusted hub.
  \item \textbf{Relayer Network.} We implement a collaborative Relayer Network that enables preference policies (e.g., fee and latency minimization) with cryptoeconomic guarantees, while allowing horizontal scaling as relayers specialize in specific chains.
  \item \textbf{Evaluation against hub-and-spoke baselines.} Using the IBC topology, we evaluate connectivity, robustness to hub removal, scalability under load, and on-chain verification overhead against direct connections and a hub-and-spoke baseline (Axelar/Cosmos Hub).
\end{enumerate}

In addition to the Relayer Network, we implement two alternative routing methods that explore different points in the design space: (1) single-relayer routing, where a single relayer computes and delivers messages; and (2) zkRouter, which uses zero-knowledge proofs to verify route computation off-chain. We discuss these in the Discussion section after evaluation.

\section{Background}

\SystemName{} is built on the IBC protocol for secure message delivery between IBC-enabled blockchains. This section describes IBC and its components, then outlines how the network has evolved. We highlight current limitations that motivate our work.

\subsection{Inter-Blockchain Communication (IBC)}
The Inter-Blockchain Communication (IBC) protocol provides secure message passing and token transfer between blockchains in a trust-minimized way. Unlike centralized bridges, IBC uses light-client verification on-chain. Each chain running IBC maintains a light client of the other chain's state. When one chain sends a packet, it is accompanied by a cryptographic proof (a Merkle proof of inclusion in that chain's state). The receiving chain uses its light client to verify the proof against the known header of the sending chain. If valid, the packet is processed. The only trust assumption is the honest majority of each chain's validators.

\subsection{Relayers}
In IBC, messages and consensus states are delivered across chains by third party relayers. Relayers monitor blockchain networks for IBC transactions, gathering the necessary data and cryptographic proofs for the receiving chain to validate and execute the transactions. 
Due to the operational costs to both run the relayer and perform on-chain operations, relayers need to be rewarded with fees for the transactions they relay. 
ICS-29~\cite{cosmos_ibc_fee} is a new standard that reimburses relayers with fees that are paid by users on the source chain. Before the existence of ICS-29, to adopt IBC and ensure delivery, most of the relayers were run by foundations of blockchains providing paid service for free. However, only two out of the top eight IBC-enabled chains based on the volume of IBC transactions have implemented ICS-29.

\subsection{Multi-Hop Extension}

The current IBC protocol can only send messages between two directly connected chains. A multi-hop extension~\cite{cosmos_ibc_multihop} has been proposed that can reuse existing connections between chains to send messages between chains that are connected through extra intermediate hops.
This standard relies on the transitive nature of trust between these chains and multi-hop verification of the consensus state along the path. 

\begin{figure}[t]
    \centering
    \includegraphics[width=\linewidth]{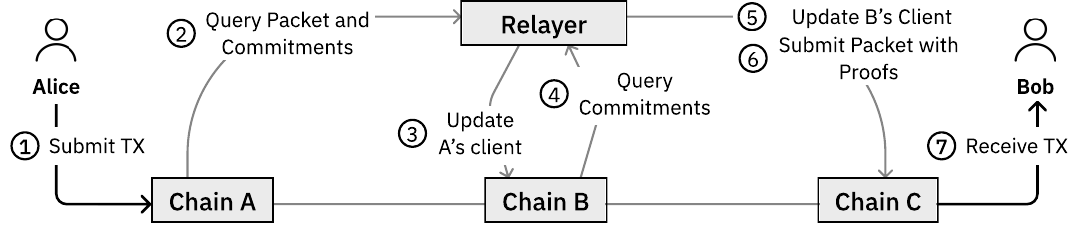}
    \caption{Sequence Diagram illustrating a multi-hop packet's flow}
    \label{fig:mhtimed}
\end{figure}

Verifying a message across multiple hops requires a chained proof showing that the first intermediate hop observed the message on the source chain, and each subsequent hop observed the previous hop. Figure~\ref{fig:mhtimed} demonstrates an example of how a cross-chain transaction from chain A to chain C is performed. First a relayer queries chain A for outgoing messages. If there are outgoing messages, it collects the current consensus state from chain A, and uses it to update the light client in chain B. It then queries chain A for the message commitment, and queries chain B for its consensus state, and delivers both to chain C, which allows chain C to verify the message originates from chain A and is correctly recorded on the source chain. 

\subsection{Limitations} \label{sec:ibc-limit}

While IBC minimizes trust assumptions, it still has practical limits. New connections require creating light clients and completing multi-step handshakes, which adds coordination and on-chain cost. This slows adoption and encourages hub-and-spoke topologies, where most chains connect to a few large hubs. Multi-hop IBC can reuse existing connections to reach more chains. However, it introduces liveness and trust dependencies on all intermediate hops and does not specify how to compute routes. Additional infrastructure is needed to select paths and enforce application policies.

\section{Architecture} \label{sec:arch}

\begin{figure*}[!ht]
    \centering
    \includegraphics[width=\textwidth]{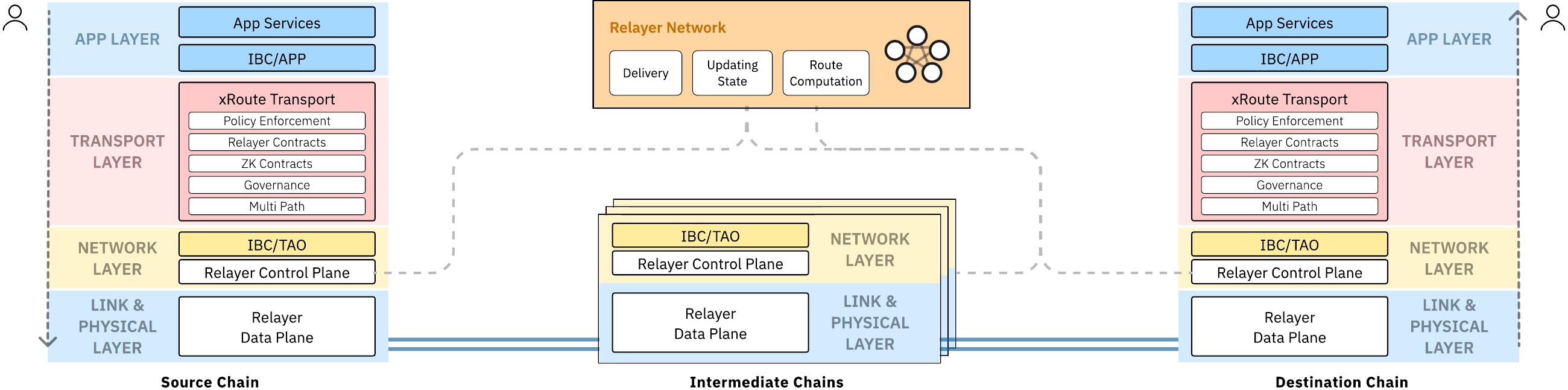}
    \caption{\SystemName{} architecture. \SystemName{} Core modules run on-chain (application services, policy enforcement, governance, and multi-path). Relayers run off-chain control/data planes for route computation and packet forwarding.}
    \label{fig:arch}
\end{figure*}

In this section, we describe the system design and architecture of \SystemName{}'s policy-driven cross-chain routing stack. Figure~\ref{fig:arch} shows the high-level architecture, organized into abstraction layers modeled after a network stack where each layer isolates a specific concern (application services, policy enforcement, route computation, and packet delivery). \SystemName{} consists of four main component layers:

\paragraph{\textbf{\SystemName{} Application Services -- \textit{App Layer}}} offers application-level services for cross-chain messaging. It enables other modules to send messages across chains seamlessly without interacting with the underlying protocols. An application can interact directly with the transport layer if it needs fine-grained control over how messages are routed and delivered across chains.

\paragraph{\textbf{\SystemName{} Transport -- \textit{Transport Layer}}} contains \SystemName{}'s core modules: (1) Policy Enforcement module enforces policies on messages and verifies integrity upon delivery. (2) Relayer Contracts provide smart contracts to facilitate the operations of the Relayer Network, such as verifying messages and fee distribution. (3) Governance module determines channels and chain identifier mappings through on-chain creation and voting of proposals. (4) Multi-Path module provides additional security guarantees for sensitive cross-chain transactions that require stronger protection against possible intermediate chain breaches.

\paragraph{\textbf{Relayer Control Plane (RCP) -- \textit{Network Layer}}}
is responsible for communicating with other relayers on the Relayer Network, and receiving computed routes from them. Routing decisions are made based on security policies, gas costs, congestion, and gas token availability of the participating relayers. 

\paragraph{\textbf{Relayer Data Plane (RDP) -- \textit{Link and Physical Layers}}} is responsible for watching the chain for new outgoing IBC packets, updating light-client states on intermediate chains, and submitting packets to the destination. These operations are done by using RPCs provided by chains.

Complementing the stack components, \SystemName{} introduces an off-chain coordination layer called the Relayer Network. This enables relayers to collaborate on route computation and message delivery without introducing additional on-chain trust assumptions; security policies remain verified solely by source and destination chains. By moving the computation off-chain and verifying on-chain, \SystemName{} supports flexible policies that would be prohibitively expensive to enforce on-chain, such as fee minimization or DEX selection.

In the following subsections, we will describe the functionality of the different modules in turn. 

\subsection{Policy Enforcement Module}

A \textbf{policy} specifies conditions that a route must satisfy (e.g., every intermediate hop must have Nakamoto coefficient $\ge 8$). We distinguish between \textbf{security policies}, which must hold for packet acceptance and are enforced deterministically on-chain via IBC that uses auxiliary data verified against the light client state (e.g., an intermediate chain's validator set, to compute Nakamoto coefficient), and \textbf{preference policies}, which should hold for best user experience (e.g., fee minimization) where enforcement depends on the routing method.

The policy module has two functions. First, it allows a user to store policies on their preferred blockchain. For instance, a user might want to specify a policy requiring all of its messages to hop through chains with specified security requirement (e.g. chain with Nakamoto coefficient greater or equal to $6$). Second, when receiving a packet, the policy module verifies if the received packet's policies match those that are expected on the destination chain. This allows for routing based on policies and ensures that packets comply with specified delivery requirements.

Policies can be of many types depending on the needs of users, including but not limited to: Security Requirements, Message Delivery Time, and Message Delivery Fees. We describe each in turn:

\subsubsection{Security Requirements}
In our design, security requirements are specified as policies and are strictly verified by the target chain. If a relayer tries to deliver a message through a route that doesn't satisfy one of the security conditions, it will be rejected.

Two examples of possible security policies are: (1) Minimum 
acceptable Nakamoto Coefficient and (2) Minimum number of validators.

\subsubsection{Message Delivery Fees}

A user can specify maximum amount of fee they are willing to pay, or may request to minimize the fees. 
While minimization is not directly verifiable on-chain (though achievable with zero-knowledge proofs), it can be achieved through economic incentives. Relayers must post a stake to participate in the relayer network; submitting a non-optimal route enables other members to slash the misbehaving relayer's stake. To guard against network collusion, an on-chain escrow mechanism on the source chain refunds a portion of the fee if another relayer (or separate relayer network) submits a cheaper route satisfying the user's requirements. This design ensures relayers compete to find efficient routes and is similar to incentive structures used in money markets to encourage liquidation.

\subsubsection{Message Delivery Time}
Two different policies can be defined for message delivery time: (1) Timeout policy, in which user can specify a deadline based on real clock time and (2) Minimum Time policy, where a user specifies that they prefer shorter routes.

\subsubsection{Policy Definition}
In IBC, channel versions can be used to encode information about specific use cases or channel properties and all endpoints are required to agree on the channel version during channel setup, such as ICS-20 channels dedicated for inter-chain token standard.
We use channel versions to specify the routing method and the security policies that channel requires. To ensure longevity of channels, additional policies are typically excluded from versioning to allow efficient reusing the channel for future messages. For example, \texttt{ics20-1/validators:4} denotes a version for an ICS-20 token transfer channel that requires at least 4 validators on intermediate hops.

\subsubsection{Policy Enforcement}
Security policies are enforced on-chain by requiring relayers to submit proofs verifying intermediate chain properties. We walk through a concrete example.

Consider Alice sending a token from Chain $A$ to Chain $C$ via Chain $B$, with a policy requiring all intermediate chains to have Nakamoto coefficient greater than 8. When a relayer updates Chain $B$'s light client on Chain $C$, it stores a consensus state $\text{CS}_C(B, h)$ containing (among other fields) a hash of Chain $B$'s validator set at the next block height $h+1$. This cryptographic commitment binds Chain $C$ to Chain $B$'s future validator set.

To deliver Alice's message at height $h$, a relayer must prove that Chain B's actual validator set satisfies the policy. The relayer queries Chain $B$ for its validator set $\mathcal{V}_{h+1}$ at height $h+1$, then submits both the standard IBC proofs and $\mathcal{V}_{h+1}$ to Chain $C$. Chain $C$ verifies two conditions: (1) the hash of $\mathcal{V}_{h+1}$ matches the committed hash in $\text{CS}_C(B, h)$, ensuring the validator set is authentic; and (2) $\mathcal{V}_{h+1}$ satisfies the Nakamoto coefficient requirement. Only if both checks succeed does Chain $C$ accept the message.

This design ensures that policy violations are detectable and enforceable on-chain, with security based on the light client's cryptographic commitments rather than trust in relayers.

\subsection{Multi-Path Module}
The multi-path module provides enhanced security by providing disjoint path transmission for transactions between the source chain and the destination chain. A user can request a transaction be sent along $N$ disjoint paths (as determined by the user as a policy) between the source and the destination chains. The user can also specify a threshold $M$ such that when $M$ successful receipts are verified, the transaction is considered complete. If any of the intermediate chains are malicious or adversarial, the altered transaction would be aborted as the multi-path module on the destination requires responses from other chains before it proceeds.

Multi-path message transmission proceeds as follows: the Application module sends message $m_{h}$ at block height $h$ on channel $C$, at which point the Multi-path middleware intercepts the message and identifies all channels $C^{\prime}$ with the same destination endpoint as $C$. The message is then sent on all channels in $C^{\prime}$. On the destination side, the Multi-path middleware acknowledges receipt and increments a counter for each received message $m_{h}$; once the number of received messages surpasses $M$, the destination middleware submits the message to the Application module and marks $m_{h}$ as received.

\subsection{Governance Module}
IBC relies on identifiers such as channel IDs and client IDs rather than counter-party chain identifier, a unique name that a chain uses to represent itself. This can be inconvenient for users, who are required to specify the channel ID to send a message. To make cross-chain communication more intuitive for end-users, \SystemName{} abstracts away these technicalities. Users only need to specify the target chain identifier in order to communicate with that chain. Since chains themselves define their own unique identifier, there is an inherent need for a mechanism to map chain identifiers onto the other identifiers used internally by IBC.

\SystemName{} supports an on-chain governance module for users to propose mappings between corresponding identifiers. Those proposals are broadcasted to all validators who then vote upon them and once validated by a defined quorum of validators, they are committed on-chain for future reference and can be retrieved as proof. 

\subsection{Relayer Network} \label{sec:relayer-network}
In this section, we first discuss the necessity of a collaborative approach in relaying messages. Second, we explain how we design a decentralized system, \textbf{Relayer Network}, that allows relayers to collaborate, compute routes, and share fees while keeping the original simplicity from the user's perspective.

\subsubsection{Rationale}

Packet relaying in original IBC is designed as a Winner-Takes-All approach and between two competing relayers, only one can receive rewards. The Multi-Hop extension, while introducing several new responsibilities, assumes all of the actions to be handled by a single relayer. These assumptions were made due to relayers being typically maintained by Blockchain Foundations. As a result of economic motivation by recently introduced fee incentives module \cite{cosmos_ibc_fee}, we expect multiple relayers to join and compete for relaying packets. The current approach is problematic for two reasons: (1) Relayer is required to monitor several chains and also hold native currencies for each chain to submit transactions, which is impractical at scale. (2) based on experiments in \cite{ibc-performance}, competing relayers could worsen transaction throughput of the chains. These reasons necessitate a better organized design to take advantage of multiple relayers being available in the system contributing to the performance, rather than wasting resources in a Winner-takes-all situation.

\subsubsection{New Relayer Roles}
We discuss new introduced relayer roles in this section:

\paragraph{Multi-Hop Channel Creators}
These entities are responsible for creating Multi-Hop channels based on computed routes between source and destination chains. They perform IBC handshake to create a new channel between two chains.

\paragraph{State Update Relayers}
To be able to pass the packet along the intermediary chains, a multi-hop proof of each chain's consensus state is needed to be queried. However, before that, each blockchain's light client state needs to be updated to reflect the new transaction on the source chain. This is the responsibility of the Update Relayer to update the consensus state of the light clients in the path in proper order.

\begin{figure*}[t]
    \centering
    \includegraphics[width=\textwidth]{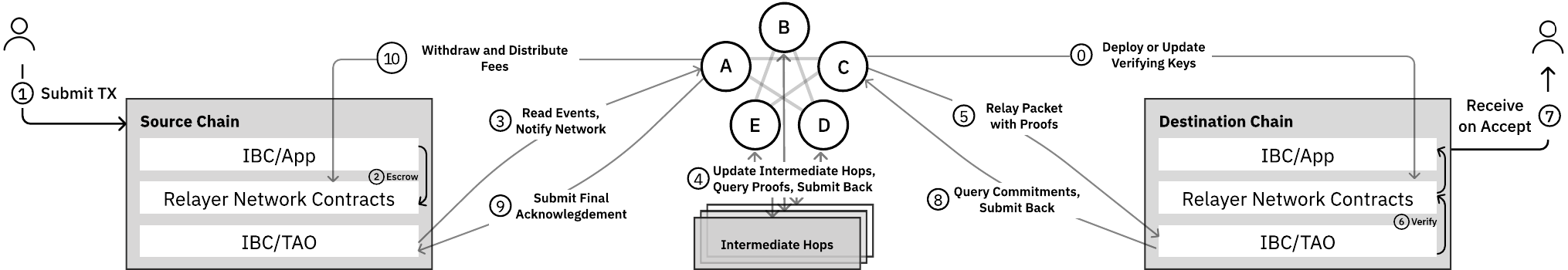}
    \caption{Relayer Network Architecture. The coordination network is a BFT blockchain where relayers stake tokens and collaborate on route computation and packet delivery. Management contracts on source/destination chains hold escrowed fees and verify relayer signatures. Relayers specialize in specific chains, each monitoring only the chains they stake for and holding gas tokens only for those chains, reducing per-relayer operational burden.}
    \label{fig:relarch}
\end{figure*}

\paragraph{Multi-Hop Packet Relayers}
These are the key relayers that are responsible for handling the user packets. They watch over the source and destination chain, getting queued messages, computing the consensus state and \SystemName{} policy proofs and finally submitting them to the destination chain. After that they will watch over the inclusion of the executed transaction or exclusion in case of failure in the destination chain and relay an acknowledgment or timeout packet back to the source chain completing the flow of a packet.

\subsubsection{Design}
Our relayer network is a co-operative network of relayers that collaborate to perform tasks more efficiently than they could independently. By splitting up these responsibilities and sharing them across multiple relayers we avoid the scalability issues described earlier and allow competing parties to share in profits proportionally based on their contribution rather than winner-takes-all. We do this by creating two core components:

\paragraph{Coordination Network}
A permissionless Byzantine fault-tolerant blockchain that is built with CosmosSDK. Relayers need to stake a minimum amount of tokens to enter this network and they face penalties upon malicious acts or unavailability. Each relayer actively collects the events from the blockchains it participates and broadcasts them to the network. Upon noticing an event for chain of interest, a relayer picks the message, does the necessary actions for it, and submits the result along with the proofs back to the network. The next relayer will notice the progress and continue the work. This process will repeat until the message is delivered and the acknowledgment packet is relayed back to the source chain. Relayer Network also maintains the current state of the Cosmos network to provide route computation and upon creation of new channels and chains, the network collectively reaches quorum on updating the chain and channel mappings.

\paragraph{Management Contract}
Management contracts reside on any chain that is either a source chain or a destination chain or both. Intermediate chains do not need to host this contract. The contract is meant to be deployed by participants of relayer network. This contract (1) is the receiver of escrowed fees from users (2) holds a list of public keys that can be updated by quorum from relayer network (3) is used to verify every message on destination to make sure it has correct routing information which is approved by relayer network.

\paragraph{Route Computation}
Each relayer in the network monitors a specific set of chains, maintaining up-to-date state through periodic RPC queries or, ideally, by running a dedicated RPC node or lightweight indexer. When a new cross-chain message is detected, the network divides the resulting tasks and assigns them to appropriate relayers. All relayers contribute current state information for chains relevant to the message, ensuring the relayer responsible for route computation has complete and fresh data. Any misbehavior including unavailability, providing stale or incorrect information, or selecting routes that violate preference policies can result in slashing of the relayer's stake by the network. This strongly disincentives against such actions.

Figure~\ref{fig:relarch} illustrates the architecture of Relayer network and full flow of a packet collaboratively routed through it.

\subsection{Multi-Chain Decentralized Applications}
\label{sec:app}

There are numerous decentralized applications that can benefit from enhanced cross-chain connectivity, especially in DeFi sector that performance significantly relies on liquidity availability. We introduce the concept of Multi-Chain dApps, which consists of smart contracts or modules that utilize \SystemName{} to further optimize their performance. On any \SystemName{}-activated blockchain, apps can access \SystemName{} interface to perform cross-chain operations seamlessly. 

\subsubsection{Example Use-case: Multi-Chain StableSwap}
We showcase a stable-swap application which is based on existing 
Constant Product Market Making (CPMM) Decentralized Exchanges and is able to use cross-chain liquidity through \SystemName{}. CPMM exchanges work based on specific curve ($x*y = k$) to determine price based on demand and liquidity available. The main problem with CPMM exchanges is high slippage, fluctuations of price, when the amount of trade is significant with respect to pool liquidity.

The flow of a swap through multi-chain StableSwap looks like: (1) User submits a stable swap request on source chain, indicating the maximum slippage that they can afford, (2) The Stable-Swap protocol provides a decision function that is available for relayers to query, (3) Relayer(s) compute a swap route, which satisfies the conditions and submit it to the source chain, triggering many multi-hop transactions. (4) Based on the available liquidity, the trade may fully or partially succeed and user gets their funds back. Multi-chain routing lets StableSwap draw on liquidity across multiple chains, reducing slippage for large trades compared to any single-chain pool. In our evaluation (Section~\ref{sec:swapeval}), 3-hop routing approaches the unified-liquidity baseline, while 0-hop swaps saturate early due to limited local liquidity.

\section{Decentralization and Security}
\label{sec:decentralization-security}

Relayers in IBC introduce a point of centralization as they are needed to deliver messages and proofs to the destination. This is acceptable because even malicious relayers cannot deviate from user-specified security policies or forge packets. The Relayer network improves upon this by distributing tasks across different relayers and using a slashing mechanism for misbehavior.

By creating a CometBFT-based blockchain for the Relayer Network, the network is safe under $f < \frac{n}{3}$ Byzantine validators in a partially synchronous network model~\cite{10.1145/42282.42283}. Participation is permissionless: any relayer may join by posting a bond on-chain.

Adversarial relayers may influence routing for fees but cannot violate security policies or alter the packet. Censorship or intentionally suboptimal routing is detectable and punishable through slashing. If an adversary compromises enough relayers to evade slashing, our on-chain fee escrow mechanism can disincentivize such attacks. The fee that the user pays for the multi-hop message is held in escrow on the source chain for a certain period, giving a third party a chance to submit evidence of misbehavior and claim the escrowed fee as a reward.

The IBC multi-hop specification does not constrain the security of intermediate hops. If an adversary compromises an intermediate chain, it can forge transaction on a source chain, forge proofs, or falsify packet information. This attack is specific to multi-hop IBC; in our system, it is possible if a chain that satisfies the security policies is compromised or malicious. However, this attack is unlikely, because it can only be done by chains that have high trust score and are rarely malicious. To mitigate this attack, we provide a multi-path module that sends messages across multiple independent routes and accepts the result only when a majority agree, providing tolerance against compromised intermediate chains.
\section{Implementation}
\label{sec:implementation}

We implemented a modular prototype consisting of: the Relayer Network (Cosmos SDK + CosmWasm), policy enforcement modules, cooperation-aware relayers, and evaluation tooling.

The Relayer Network is built as a Cosmos SDK chain with CosmWasm contracts providing upgradeability. Core contracts contain the logic for relayer lifecycle management (joining and leaving the network), governance (updating chain and client identifiers), accounting (fee withdrawals), and slashing conditions. Management contracts are also deployed on the source and destination chains, enabling users to escrow fees, relayers to collect fees, and destination to verify routing decision authenticity.
Our Cosmos SDK modules provide a policy enforcement interface accepting decisions from the Relayer Network. Integrating this module makes any chain \SystemName{}-compatible.

For cooperative relaying, we implemented a Rust-based relayer with specialization (relayers declare their chains of interest) and path scheduling. We 
identified that excessive client updates dominated multi-hop latency, and addressed this with a minimal-update algorithm that tracks update dependencies across hops and batches updates where possible, significantly reducing unnecessary traffic.
For evaluation, we developed a multi-hop load generator that measures throughput and latency across various network topologies, along with a SimPy~\cite{simpy}-based simulation model for congestion analysis. All components described in this section are open sourced and available at \texttt{https://github.com/AminRezaei0x443/xroute}.

\section{Evaluation}
Our evaluation compares our system with a hub-and-spoke cross-chain message-passing system, the predominant approach for providing connectivity among Cosmos chains. We measure connectivity, decentralization, and scalability; evaluate required on-chain transaction fees and off-chain maintenance costs; and benchmark system components. Finally, we analyze a DeFi use case to quantify the impact on asset liquidity when a protocol can trade tokens seamlessly across multiple chains.

\subsection{Methodology}
For our experiments, we examined the current network topology and effect of different parameters on several essential metrics: connectivity, decentralization, scalability, and costs. We describe our data collection and simulation methodology below:

\paragraph{Data Collection}
We collected Cosmos network topology from chain registry~\cite{cosmos_chain_registry} via RPC (connections, channels, clients). All the data are up to Apr. 2025 and correspond to \texttt{addf89a9} commit on the chain registry repository.

\paragraph{Simulation}
To evaluate the scalability of our system and also StableSwap's performance, we ran simulations using Simpy~\cite{simpy}, due to the prohibitively high cost of running large-scale experiments on real blockchains.

\subsection{Connectivity}

As mentioned in Section~\ref{sec:ibc-limit}, direct IBC connections requires manual setup, limiting connectivity.
Therefore, one of the key metrics that we want to measure is the degree of additional connectivity that \SystemName{}'s multi-hop IBC approach provides over that of single-hop IBC. We also compare \SystemName{} with Axelar, which represents a hub-and-spoke model where the Axelar chain serves as a connecting hub for other chains. For this analysis, we assume the direct IBC topology is fixed, and make use of the IBC topology that we collected in April 2025.

Additionally, for multi-hop IBC and Axelar, intermediate hops must meet minimum Nakamoto coefficient~\cite{srinivasan2017quantifying}. The blockchain community lacks a standardized metric for measuring decentralization and trust. Although the Nakamoto coefficient's validity as a security measure is debated, it is straightforward to compute for any public chain and suffices for this comparative analysis. Alternative metrics could be integrated into \SystemName{}, such as total staked value in USD, which may better reflect chain security at the cost of requiring a price oracle. For Axelar, because it always serves as the intermediate hop, we apply the same security threshold to the Axelar chain itself.

\begin{figure}[t]
    \centering
    \includegraphics[width=0.45\textwidth]{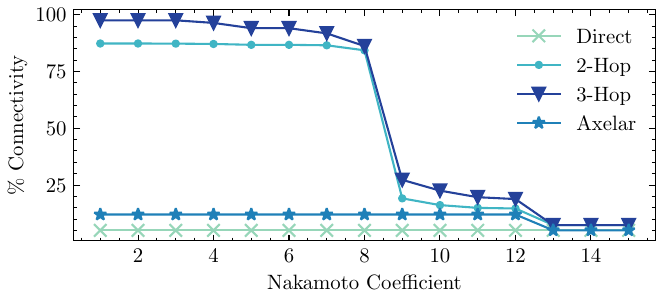}
    \caption{Effect of utilizing \SystemName{} on Connectivity, when chains with smaller than required Nakamoto coefficient are avoided}
    \label{fig:nks}
\end{figure}

Figure~\ref{fig:nks} shows that we can provide nearly 90\% connectivity between chains while only allowing chains with a Nakamoto coefficient of 8 or higher to serve as intermediate hops. This compares to less than 15\% connectivity for Axelar, which is limited by the number of direct IBC connections that Axelar chain has to other chains, and approximately 5\% connectivity when relying only on direct connections. Connectivity drops precipitously at a Nakamoto coefficient of 9 and higher as that is a higher than the coefficient of some of the largest, most strongly connected Cosmos chains. These results illustrate the need for multi-hop message passing to realize the goal of providing connectivity between most chains.

\subsection{Decentralization}

Decentralization is one of the key defining properties of blockchains. Not only does decentralization offer stronger security compared to centralized approaches, it also provides higher availability as a chain can continue to function even if a minority of validators become unavailable. \SystemName{}'s multi-hop routing, where any chain that meets a channel's security policy requirements can serve as an intermediate hop, offers a higher degree of decentralization compared to hub-and-spoke approaches where the failure of the hub chain can cause connectivity to revert back to the level offered by using only direct IBC connections.

\begin{figure}[t]
    \centering
    \includegraphics[width=0.45\textwidth]{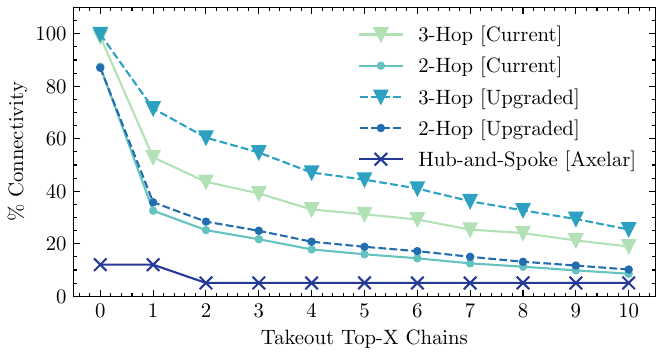}
    \caption{Effect of Taking out Top-X chains (sorted by number of connections) on connectivity.}
    \label{fig:takeout}
\end{figure}

To measure \SystemName{}'s impact on decentralization, we perform an experiment where we measure the change in chain connectivity as we remove the most strongly connected chains, simulating the worst case scenario where those large chains become unavailable. Figure~\ref{fig:takeout} shows \SystemName{}'s result for 2-hop and 3-hop routing using both the current IBC topology, and an upgraded topology with additional connectivity between smaller chains. 

For the current IBC topology, the loss of just the most strongly connected chain reduces connectivity to less than $40\%$. This is because most small chains are only connected to the largest chains, and the connectivity between small chains can be disrupted when a large chain fails even with multi-hop routing. \SystemName{} improves decentralization by enabling paths longer than 2 hops and provides better connectivity and reliability, even in the event of large chain failures.

One approach to further improve decentralization is to incentivize the creation of direct IBC connections between smaller chains. In \SystemName{}, direct connections between two small chains with low Nakamoto coefficients can be bootstrapped using a multi-hop IBC connection with a high Nakamoto coefficient intermediate chain without requiring manual intervention by the chains' stakeholders/developers. Note that some applications may choose to not use connections bootstrapped in this way, as it relies on trusting the past security of a chain rather than its current security. In this analysis, we only consider applications that accept multi-hop bootstrapping of direct connections.

The upgraded lines in Figure~\ref{fig:takeout} show the increase in decentralization by incentivizing each chain to create one additional direct connection to a random chain that it was not previously connected to with a Nakamoto coefficient of at least 6. Using 3-hop routing, \SystemName{} can still provide nearly 80\% connectivity between chains when the most strongly connected chain becomes unavailable. In the unlikely event that the four most strongly connected chains become unavailable, \SystemName{} can still provide nearly 50\% connectivity between chains using 3-hop routing.

Note that we do not measure the degree of decentralization of the relayer network. Although the relayer network serves an important purpose in the \SystemName{} framework, the damage that can be caused by malicious behavior from a compromised relayer network is primarily limited to denial of service and higher fees as security properties can be verified by the source and destination chains.

\subsection{Scalability}

Current hub-and-spoke based approaches require that messages are first sent from the source chain to the hub chain, and then sent again from the hub chain to the destination chain. This can limit scalability as the maximum rate of cross chain messages is bounded by the throughput of the hub chain. Most Cosmos chains, including Axelar, are built using the CosmosSDK~\cite{cosmos-sdk} and use the Tendermint consensus algorithm~\cite{buchman2016tendermint}. Results from the original Tendermint whitepaper~\cite{buchman2016tendermint} show that it can sustain approximately 4000 transactions per second for a 64-node deployment. Therefore, the hub chain will experience high congestion once its throughput is saturated by cross-chain messages, which can result in high fees for hub chains with working fee markets, or failed transactions for those that do not.

Unlike hub-and-spoke approaches, \SystemName{} does not need to send messages to intermediate chains. Rather, it only needs to update the consensus state of intermediate chains to generate multi-hop proofs. Furthermore, instead of a single hub chain, different multi-hop paths can be used for delivering cross-chain message. This spreads the load across chains that meet the security policy requirements.

\begin{figure}[t]
    \centering
    \includegraphics[width=0.45\textwidth]{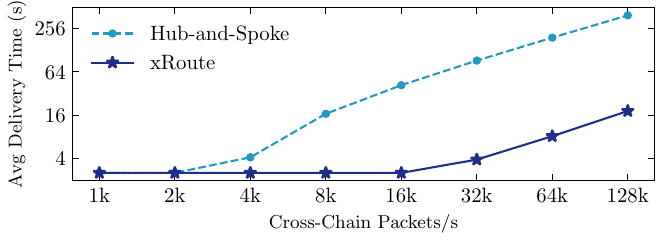}
    \caption{Packet Delivery Time in Hub-and-Spoke compared to \SystemName{}}
    \label{fig:scale}
\end{figure}

To evaluate scaling, we simulate a network of 500 blockchains connected to a single hub and send cross-chain messages through it at varying rates. Packets are generated and assigned to paths based on a Zipf distribution. We measure the average end-to-end time from message inclusion on the source chain to receipt on the destination chain, comparing hub-and-spoke and \SystemName{}. Figure~\ref{fig:scale} shows the results: \SystemName{} handles up to 32k packets per second without latency impact compared to 4k for hub-and-spoke, and tolerates higher loads with reasonable latency. Congestion can be further mitigated in \SystemName{} if fee-minimizing routes are selected as applications will prefer to use low fee channels that avoid highly congested chains.

\subsection{Costs}

In this section, we compare the gas cost between sending a multi-hop message using \SystemName{} and sending the same message manually using one or more direct IBC connections through the same intermediate chains. \SystemName{} has an initial setup cost for establishing a multi-hop channel for a source/destination pair. However, the created channel is usually long-lived and reused, so we amortized the cost of establishing the channel over multiple messages.

\begin{figure}[t]
    \centering
    \includegraphics[width=0.45\textwidth]{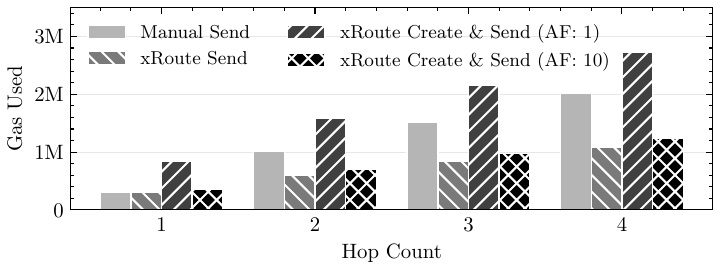}
    \caption{Gas costs of \SystemName{} vs Manual Multi-Hop}
    \label{fig:costs}
\end{figure}

Figure~\ref{fig:costs} shows the cost of manually sending the message using direct IBC connections, using \SystemName{} without including the cost of establishing the channel, and using \SystemName{} with the cost of establishing the channel with an amortization factor of 1 message and 10 messages. As expected, the cost increases with higher hop count, but the rate of increase for \SystemName{} is lower than manually sending the messages if the cost of establishing a channel is either not included or amortized.

We also measure the cost of performing on-chain verification of additional security policies. For this experiment, we send 12 messages along a 2-hop channel where the intermediate chain has 4 validators.
We compare the cost of sending these messages against the cost of sending the same messages without the security policies.
Since the gas cost depends entirely on the amount of data being processed, we found that the average additional gas cost per validator per hop is $15459$ gas units.  For the current state of the Cosmos network, where the network diameter is $3$ and average number of validators is $54$, this would result in $15459 * 3 * 54 = 2504358$ gas units which would be less than $0.05$ USD using current price of ATOM, the gas token for many Cosmos chains.
These results show that our approach is cost effective, with fees that are low enough such that users will unlikely choose to perform manual single-hop transfers for fee-related reasons.

\subsection{Usecase: Stable Swap} \label{sec:swapeval}

As described in Section~\ref{sec:app}, we implemented a token swap protocol that breaks down a large swap into multiple smaller swaps performed on other chains using \SystemName{}. The protocol takes advantage of available liquidity across multiple chains to provide the user with a better price for their swap.
This experiment compares the financial gains of our swap protocol when trading USDT for USDC against different baselines, where USDC and USDT are both stablecoins pegged to USD. We collected total available liquidity on USDT-USDC pools across Cosmos in January 2025 and scaled it proportionally to 10 billion USD to demonstrate the effect of large trades.

\begin{figure}[t]
    \centering
    \includegraphics[width=0.45\textwidth]{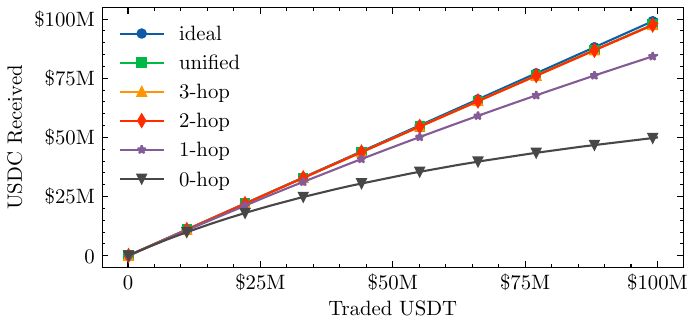}
    \caption{StableSwap performance between USDT and USDC}
    \label{fig:swap}
\end{figure}

Figure \ref{fig:swap} shows the amount of USDC received from trading a given amount of USDT using the different approaches. The ideal line illustrates perfect 1:1 pricing, which is the expected ratio from trading two stablecoins pegged to the same currency. The unified line illustrates the  case where all available liquidity across all Cosmos chains is somehow unified to a single chain. The results are very close to the ideal line even when trading \$100 million USDT tokens.

The 0-hop line illustrates the case where the trade is satisfied only using the available liquidity of the local chain. For this experiment, we selected the Cosmos chain with the most available liquidity as the local chain to provide the best case scenario for the 0-hop approach. Trading \$100 million USDT would return less than \$50 million USDC because of the limited liquidity available in the local chain. Going from 0-hop to 1-hop significantly increases the available liquidity and the return in USDC from trading USDT. The 3-hop results are nearly identical to the unified results.

One might argue that such large trades are not commonly performed on decentralized exchanges. However, larger trades and higher trade volumes often occur during significant market volatility. For example, decentralized exchanges experienced all-time high trade volumes on March 11, 2023 at \$20 billion USD during the USDC depeg event as USDC holders exchanged their USDC for other stablecoins even when there was limited available liquidity~\cite{Ledger2023}.

\section{Discussion}
Beyond the Relayer Network-based design, we explored alternative route computation approaches, each with distinct tradeoffs. We discuss these below.

\paragraph{Single Relayer Routing}
For any given message, one relayer will handle all of its necessary jobs, from computing the route to delivering the packet and submitting back the acknowledgment packet. The relayer periodically queries blockchain RPCs to maintain an up-to-date network view, then computes routes satisfying the message's policies upon observing a new message. The primary advantage is simplicity, but preference policies are provided based on pure incentivization. Also, this method is challenging because the relayer needs to monitor all the chains on the path and hold gas currencies for all of them.

\paragraph{Zero-Knowledge Routing}
This method computes routes off-chain and provides on-chain verifiable correctness via zero-knowledge proofs. Our implementation uses RISC Zero's zkVM
\cite{risczero2024}: (1) the program gets the necessary info to compute the route, which consists of the message with policies and the latest state of the network queried from RPCs along with the Merkle proofs, (2) verifies the inputs, (3) computes and outputs the route along with the block header that these inputs corresponded to. Every time the program is executed, it also generates a compact proof of correctness. The benefits of this approach are its flexibility to adapt different policies and providing optimality guarantees. The verification process is inexpensive, ensuring low overhead for chain operators. However, generating these proofs demands significant computational resources (GPUs).

\paragraph{Link-State Routing}
We implemented link-state routing by storing routing tables on-chain, enabling calculation of the next hop for outgoing messages. While straightforward to build, this approach incurred significant gas costs for querying routes and updating tables due to on-chain read and write operations, making it impractical for real-world scenarios.

\section{Related Work}
Cross-chain communication has received significant attention due to financial implications; compromises can lead to millions in losses. Surveys categorize approaches~\cite{10646648, 10664224, 10.1145/3471140, r3_chain_interoperability_2016, 10.1145/3648607, 10.1145/3573896, https://doi.org/10.1049/blc2.12032, 10.1145/3333165.3333167}.

\paragraph{Notary-Based Mechanisms}
A number of approaches rely on Notaries, systems that listen to the source chain for events and then confirm the events on the target chain. These solutions involve different degrees of trust assumptions, existing from central entities, to multi-sig wallets and decentralized methods.
Chainlink's Cross-Chain Interoperability Protocol (CCIP) \cite{ccip} is built upon Decentralized Oracle Networks (DONs) acting as Notaries, but also relies on its Risk Management Network to detect malicious behavior in these DONs.
Hyperlane \cite{hyperlane} and LayerZero \cite{layer-zero-wp, layerzero_network} also utilize the same concept by introducing more smart contracts, the former defines flexible inter-chain security modules (ISMs) while the latter uses decentralized verifier networks (DVNs) which provides better trust assumptions.  Omnity (previously known as Octopus) \cite{omnity_network, omnity_cross_chain_endgame} deploys its main logic on ICP (Internet Computer) blockchain that uses its own unique cryptographic primitive known as Chain-Key Cryptography \cite{internetcomputer_chain_key_cryptography}, a threshold signature method that can be created for any contract data by validators, to confirm events as simple as a signature verification.
This model is also widely adopted by many bridges which use multi-signature wallets to control the notary and upon breach lead to million dollar loss. While decentralized solutions provide enhanced security, this approach still needs the additional trust assumption on the notary.

\paragraph{Native Interoperability and Hub-and-Spoke Architectures}
The most famous example of this approach is the Cosmos ecosystem
\cite{cosmos-wp}, which uses its IBC protocol and Light Clients for verification to connect chains through a hub. Because of native interoperability based on Light Clients, it comes with minimized trust assumptions.
Axelar \cite{axelar-wp} is a blockchain built on top of Cosmos SDK that runs as a zone within the larger Cosmos ecosystem. The chain itself provides abstraction known as gateways, which can be used to transfer tokens and execute smart contract calls across ecosystems in a path-agnostic manner.
Polkadot \cite{polkadot-wp} is another hub-and-spoke solution that introduces its own Relay Chain that allows developers to build parachains (i.e., side-chains that are not fully independent). In order for a new parachain to join the network, it must participate in slot auction and lock enough DOT tokens to lease the slot. The reliance on a single hub for governance, security, and message passing limits scalability along with the additional trust assumption.
In this class of approaches, a new blockchain acts as the "router" for transactions between ecosystems
Wang et al. \cite{10.1145/3070617.3070634} present a design that resembles Cosmos, but emphasizes crypto-economic security over trustless relayers and light clients. Anlink \cite{anlink}, an extension of this work, continues the work by enhanced scalability.
Ding et al.'s InterChain \cite{Ding2018InterChainA} focuses on cross-chain asset transaction but requires some Sub-chain nodes to be connected to their router, introducing compatibility issues and also certain trust assumptions. Kan et al. \cite{8431965} design a multi-layer system and implement a three-phase commit protocol across chains.
Limitations of these approaches include introducing additional trust assumptions through the router blockchain and a lack of practical implementation.

\paragraph{Smart Contract-Based Solutions} Implement the cross-chain transfer logic deployed on the source chain's and destination chain's smart contracts. There are two major methods in this category: (1) systems that introduce a trust assumption on third-party entity/chain to verify transfers \cite{pupyshev2020susyblockchainagnosticcrosschainasset} (2) light client implementations in smart contracts that rely on trustless relayers \cite{10.1145/3319535.3355503, 10.1145/3548606.3560652, stone2021trustlessprivacypreservingblockchainbridges}. Although the latter approach has minimized trust assumptions, they usually have high computational requirements and costs, especially if the source and destination blockchains are heterogeneous.

\paragraph{Asset Transfer-Focused Solutions}
A number of systems have focused on facilitating asset transfers and exchanges across blockchains. Some implement similar techniques to what we described above \cite{0x_org, 0x_white_paper, hardjono2021blockchaingatewaysbridgesdelegated, 11183779}, while the others use cryptographic primitives like Hash Time Locking Agreements (HTLAs) \cite{ark_io}. These designs are inherently limited by their inability to support other use cases such as arbitrary message passing or smart contract execution across ecosystems.

\section{Conclusions}

In this paper, we described \SystemName{}, a secure routing and message delivery framework that significantly increases the connectivity between IBC-enabled blockchains through multi-hop IBC. \SystemName{} uses relayers to deliver cross-chain messages to the destination chain, and offloads route computation to \SystemName{}'s relayer network, which generates routes offchain that satisfy application-specified security policies and optimizes for a performance metric.

Our evaluation results show that \SystemName{} can provide more than 80\% connectivity between chains even when intermediate chains must have a Nakamoto coefficient of 8 and higher, compared to less than 20\% when using Axelar, a hub-and-spoke-based approach. Our results also show that \SystemName{} offers significantly higher decentralization and scalability compared to our baselines. We believe that this work takes us one step closer to solving the inter-blockchain communication problem and hope that the eventual solution will provide blockchains the same type of transformative growth that the Internet experienced from seamlessly connecting different networks.

\bibliographystyle{IEEEtran}
\bibliography{IEEEabrv,refs}

@INPROCEEDINGS{11183779,
  author={Maurya, Siddharth and Awathare, Nitin and Ribeiro, Vinay Joseph and Bellur, Umesh},
  booktitle={2025 IEEE 45th International Conference on Distributed Computing Systems (ICDCS)}, 
  title={Tombolo: Towards a Decentralized Interconnected Blockchain Ecosystem using Cross-Chain Payment Channels}, 
  year={2025},
  volume={},
  number={},
  pages={1044-1054},
  doi={10.1109/ICDCS63083.2025.00105}}

@online{srinivasan2017quantifying,
  author = {Srinivasan, Balaji S.},
  title = {Quantifying Decentralization},
  year = {2017},
  url = {https://news.earn.com/quantifying-decentralization-e39db233c28e},
  urldate = {2023-10-25},
  publisher = {Earn.com News},
  note = {Accessed: 2023-10-25}
}

@article{10.1145/42282.42283,
author = {Dwork, Cynthia and Lynch, Nancy and Stockmeyer, Larry},
title = {Consensus in the presence of partial synchrony},
year = {1988},
issue_date = {April 1988},
publisher = {Association for Computing Machinery},
address = {New York, NY, USA},
volume = {35},
number = {2},
issn = {0004-5411},
url = {https://doi.org/10.1145/42282.42283},
doi = {10.1145/42282.42283},
journal = {J. ACM},
month = apr,
pages = {288–323},
numpages = {36}
}

@article{defillama,
  author    = {},
  title     = {{All Chains Bridged TVL - DefiLlama}},
  journal   = {https://defillama.com/bridged},
  month     = {September},
  year      = {2024},
}

@article{bridge-vulnerabilities,
  author    = {},
  title     = {{7 Cross-Chain Bridge Vulnerabilities Explained | Chainlink}},
  journal   = {https://chain.link/education-hub/cross-chain-bridge-vulnerabilities},
  month     = {September},
  year      = {2024},
}

@misc{risczero2024,
    title = {RISC Zero API Documentation},
    author = {RISC Zero},
    year = {2024},
    url = {https://dev.risczero.com/api},
    note = {Accessed: 2024-08-16}
}

@misc{cosmos_chain_registry,
    title = {Cosmos Chain Registry},
    author = {Cosmos Network},
    year = {2024},
    url = {https://github.com/cosmos/chain-registry},
    note = {Accessed: 2024-08-16}
}

@misc{cosmos_ibc_multihop,
    title = {IBC Specification: Multi-Hop Communication (ICS-033)},
    author = {Cosmos Network},
    year = {2024},
    url = {https://github.com/cosmos/ibc/blob/main/spec/core/ics-033-multi-hop/README.md},
    note = {Accessed: 2024-08-16}
}

@misc{cosmos_ibc_fee,
    title = {IBC Specification: Fee Payment (ICS-029)},
    author = {Cosmos Network},
    year = {2024},
    url = {https://github.com/cosmos/ibc/blob/main/spec/app/ics-029-fee-payment/README.md},
    note = {Accessed: 2024-12-07}
}

@misc{simpy,
	author = {},
	title = {SimPy},
	howpublished = {\url{https://gitlab.com/team-simpy/simpy/}},
	year = {},
}

@phdthesis{buchman2016tendermint,
  title={Tendermint: Byzantine fault tolerance in the age of blockchains},
  author={Buchman, Ethan},
  year={2016},
  school={University of Guelph}
}

@INPROCEEDINGS{ibc-performance,
  author={Chervinski, João Otávio and Kreutz, Diego and Xu, Xiwei and Yu, Jiangshan},
  booktitle={2023 53rd Annual IEEE/IFIP International Conference on Dependable Systems and Networks (DSN)}, 
  title={Analyzing the Performance of the Inter-Blockchain Communication Protocol}, 
  year={2023},
  volume={},
  number={},
  pages={151-164},
  doi={10.1109/DSN58367.2023.00026}
}

@unpublished{cosmos-sdk,
    author = "",
    title = "Cosmos SDK",
    note = "https://github.com/cosmos/cosmos-sdk"
}

@misc{layer-zero-wp,
    author = "Zarick, Ryan and Pellegrino, Bryan and Zhang, Isaac and Kim, Thomas and Banister, Caleb",
    year = {2024},
    month = "Jan",
    title = "LayerZero",
    howpublished = {\url{https://layerzero.network/publications/LayerZero_Whitepaper_V2.1.0.pdf}}
}

@online{Ledger2023,
  author = {Ledger Insights},
  title = {USDC depeg highlights importance of Fed to banking system},  
  note = {https://www.ledgerinsights.com/usdc-depeg-fed-federal-reserve/},
  urldate = {2023-03-15},
  publisher = {Ledger Insights},
}

@unpublished{axelar-wp,
    author = "Axelar",
    title = "Axelar: Connecting Applications with Blockchain Ecosystems",
    year = {2021},
    month = {Jan},
    note = "https://www.axelar.network/whitepaper [Accessed 22/08/2024]"
}

@misc{polkadot-wp,
    author = "Wood, Gavin",
    title = "Polkadot: Vision for a Heterogeneous Multi-Chain Framework",
    Year = {2016},
    Month = {Nov},
    howpublished = {\url{https://assets.polkadot.network/Polkadot-whitepaper.pdf}}
}

@misc{cosmos-wp,
    author = "Kwon, Jae and Buchman, Ethan",
    title = "Cosmos",
    year = {2019},
    month = {Jan},
    howpublished = {\url{https://github.com/cosmos/cosmos/blob/master/WHITEPAPER.md }},
    note = "[Accessed 22/08/2024]"
}

@unpublished{ccip,
    author = "Chainlink",
    title = "Introducing the Cross-Chain Interoperability Protocol (CCIP) for Decentralized Inter-Chain Messaging and Token Movements",
    note = "https://blog.chain.link/introducing-the-cross-chain-interoperability-protocol-ccip/ [Accessed 21/08/2024]"
}

@misc{stone2021trustlessprivacypreservingblockchainbridges,
      title={Trustless, privacy-preserving blockchain bridges}, 
      author={Drew Stone},
      year={2021},
      eprint={2102.04660},
      archivePrefix={arXiv},
      primaryClass={cs.CR},
      url={https://arxiv.org/abs/2102.04660}, 
}

@INPROCEEDINGS{10646648,
  author={Augusto, André and Belchior, Rafael and Correia, Miguel and Vasconcelos, André and Zhang, Luyao and Hardjono, Thomas},
  booktitle={2024 IEEE Symposium on Security and Privacy (SP)}, 
  title={SoK: Security and Privacy of Blockchain Interoperability}, 
  year={2024},
  volume={},
  number={},
  pages={3840-3865},
  doi={10.1109/SP54263.2024.00255}}

@INPROCEEDINGS{10664224,
  author={Subramanian, Shankar and Augusto, André and Belchior, Rafael and Vasconcelos, André and Correia, Miguel},
  booktitle={2024 IEEE International Conference on Blockchain (Blockchain)}, 
  title={Benchmarking Blockchain Bridge Aggregators}, 
  year={2024},
  volume={},
  number={},
  pages={37-45},
  doi={10.1109/Blockchain62396.2024.00015}}

@article{10.1145/3471140,
author = {Belchior, Rafael and Vasconcelos, Andr\'{e} and Guerreiro, S\'{e}rgio and Correia, Miguel},
title = {A Survey on Blockchain Interoperability: Past, Present, and Future Trends},
year = {2021},
issue_date = {November 2022},
publisher = {Association for Computing Machinery},
address = {New York, NY, USA},
volume = {54},
number = {8},
issn = {0360-0300},
url = {https://doi.org/10.1145/3471140},
doi = {10.1145/3471140},
journal = {ACM Comput. Surv.},
month = oct,
articleno = {168},
numpages = {41}
}

@misc{ark_io,
  author       = {{ARK Association}},
  title        = {ARK Official Website},
  howpublished = {\url{https://ark.io/}},
  year         = {2025},
  note         = {Accessed: 2025-01-31}
}

@techreport{r3_chain_interoperability_2016,
  author       = {Buterin, Vitalik},
  title        = {Chain Interoperability},
  institution  = {R3},
  year         = {2016},
  month        = {September},
  url          = {https://cognizium.io/uploads/resources/R3%20-%20Chain%20Interoperability%20-%202016%20-%20Sep.pdf},
  note         = {Accessed: 2025-01-31}
}

@misc{hardjono2021blockchaingatewaysbridgesdelegated,
      title={Blockchain Gateways, Bridges and Delegated Hash-Locks}, 
      author={Thomas Hardjono},
      year={2021},
      eprint={2102.03933},
      archivePrefix={arXiv},
      primaryClass={cs.CR},
      url={https://arxiv.org/abs/2102.03933}, 
}

@inproceedings{10.1145/3548606.3560652,
author = {Xie, Tiancheng and Zhang, Jiaheng and Cheng, Zerui and Zhang, Fan and Zhang, Yupeng and Jia, Yongzheng and Boneh, Dan and Song, Dawn},
title = {zkBridge: Trustless Cross-chain Bridges Made Practical},
year = {2022},
isbn = {9781450394505},
publisher = {Association for Computing Machinery},
address = {New York, NY, USA},
url = {https://doi.org/10.1145/3548606.3560652},
doi = {10.1145/3548606.3560652},
booktitle = {Proceedings of the 2022 ACM SIGSAC Conference on Computer and Communications Security},
pages = {3003–3017},
numpages = {15},
location = {Los Angeles, CA, USA},
series = {CCS '22}
}

@article{10.1145/3648607,
author = {Belchior, Rafael and S\"{u}\ss{}enguth, Jan and Feng, Qi and Hardjono, Thomas and Vasconcelos, Andr\'{e} and Correia, Miguel},
title = {A Brief History of Blockchain Interoperability},
year = {2024},
issue_date = {October 2024},
publisher = {Association for Computing Machinery},
address = {New York, NY, USA},
volume = {67},
number = {10},
issn = {0001-0782},
url = {https://doi.org/10.1145/3648607},
doi = {10.1145/3648607},
journal = {Commun. ACM},
month = sep,
pages = {62–69},
numpages = {8}
}

@article{10.1145/3573896,
author = {Han, Panpan and Yan, Zheng and Ding, Wenxiu and Fei, Shufan and Wan, Zhiguo},
title = {A Survey on Cross-chain Technologies},
year = {2023},
issue_date = {June 2023},
publisher = {Association for Computing Machinery},
address = {New York, NY, USA},
volume = {2},
number = {2},
url = {https://doi.org/10.1145/3573896},
doi = {10.1145/3573896},
journal = {Distrib. Ledger Technol.},
month = jun,
articleno = {15},
numpages = {30}
}

@article{https://doi.org/10.1049/blc2.12032,
author = {Li, Li and Wu, Jiahao and Cui, Wei},
title = {A review of blockchain cross-chain technology},
journal = {IET Blockchain},
volume = {3},
number = {3},
pages = {149-158},
doi = {https://doi.org/10.1049/blc2.12032},
url = {https://ietresearch.onlinelibrary.wiley.com/doi/abs/10.1049/blc2.12032},
eprint = {https://ietresearch.onlinelibrary.wiley.com/doi/pdf/10.1049/blc2.12032},
year = {2023}
}

@misc{pupyshev2020susyblockchainagnosticcrosschainasset,
      title={SuSy: a blockchain-agnostic cross-chain asset transfer gateway protocol based on Gravity}, 
      author={Aleksei Pupyshev and Elshan Dzhafarov and Ilya Sapranidi and Inal Kardanov and Shamil Khalilov and Sten Laureyssens},
      year={2020},
      eprint={2008.13515},
      archivePrefix={arXiv},
      primaryClass={cs.CR},
      url={https://arxiv.org/abs/2008.13515}, 
}

@inproceedings{10.1145/3319535.3355503,
author = {Liu, Zhuotao and Xiang, Yangxi and Shi, Jian and Gao, Peng and Wang, Haoyu and Xiao, Xusheng and Wen, Bihan and Hu, Yih-Chun},
title = {HyperService: Interoperability and Programmability Across Heterogeneous Blockchains},
year = {2019},
isbn = {9781450367479},
publisher = {Association for Computing Machinery},
address = {New York, NY, USA},
url = {https://doi.org/10.1145/3319535.3355503},
doi = {10.1145/3319535.3355503},
booktitle = {Proceedings of the 2019 ACM SIGSAC Conference on Computer and Communications Security},
pages = {549–566},
numpages = {18},
location = {London, United Kingdom},
series = {CCS '19}
}

@inproceedings{10.1145/3070617.3070634,
author = {Wang, Hui and Cen, Yuanyuan and Li, Xuefeng},
title = {Blockchain Router: A Cross-Chain Communication Protocol},
year = {2017},
isbn = {9781450352307},
publisher = {Association for Computing Machinery},
address = {New York, NY, USA},
url = {https://doi.org/10.1145/3070617.3070634},
doi = {10.1145/3070617.3070634},
booktitle = {Proceedings of the 6th International Conference on Informatics, Environment, Energy and Applications},
pages = {94–97},
numpages = {4},
location = {Jeju, Republic of Korea},
series = {IEEA '17}
}

@inproceedings{Ding2018InterChainA,
  title={InterChain : A Framework to Support Blockchain Interoperability},
  author={Donghui Ding},
  year={2018},
  url={https://api.semanticscholar.org/CorpusID:51991936}
}

@INPROCEEDINGS{8431965,
  author={Kan, Luo and Wei, Yu and Hafiz Muhammad, Amjad and Siyuan, Wang and Gao, Ling Chao and Kai, Hu},
  booktitle={2018 IEEE International Conference on Software Quality, Reliability and Security Companion (QRS-C)}, 
  title={A Multiple Blockchains Architecture on Inter-Blockchain Communication}, 
  year={2018},
  volume={},
  number={},
  pages={139-145},
  doi={10.1109/QRS-C.2018.00037}}

@misc{anlink,
  author       = {ZhongAn Tech.},
  title        = {Anlink Whitepaper},
  howpublished = {\url{https://alicliimg.clewm.net/049/389/1389049/1484820492640c2baf37ea3e4f9fd77bd52c2a1e9bbbe1484820484.pdf}},
  year         = {2017},
  note         = {Accessed: 2025-01-31}
}

@inproceedings{10.1145/3333165.3333167,
author = {Qasse, Ilham A. and Abu Talib, Manar and Nasir, Qassim},
title = {Inter Blockchain Communication: A Survey},
year = {2019},
isbn = {9781450360890},
publisher = {Association for Computing Machinery},
address = {New York, NY, USA},
url = {https://doi.org/10.1145/3333165.3333167},
doi = {10.1145/3333165.3333167},
booktitle = {Proceedings of the ArabWIC 6th Annual International Conference Research Track},
articleno = {2},
numpages = {6},
location = {Rabat, Morocco},
series = {ArabWIC 2019}
}

@misc{internetcomputer_chain_key_cryptography,
  author       = {DFINITY Foundation},
  title        = {Chain Key Cryptography},
  howpublished = {\url{https://internetcomputer.org/how-it-works/\#Chain-key-cryptography}},
  year         = {2025},
  note         = {Accessed: 2025-01-31}
}

@misc{omnity_network,
  author       = {Omnity Network},
  title        = {Omnity Network Official Website},
  howpublished = {\url{https://www.omnity.network/}},
  year         = {2025},
  note         = {Accessed: 2025-01-31}
}

@techreport{omnity_cross_chain_endgame,
  author       = {Omnity},
  title        = {Omnity: The Cross-Chain Endgame for a Modular Blockchain World},
  institution  = {Omnity Network},
  year         = {2025},
  month        = {January},
  url          = {https://docs.google.com/document/d/10RSrGfj2Z-LmVhuAXX0Jon-9P9CO6k5ZNfS2hrQB72I/edit?pli=1&tab=t.0#heading=h.5qlvro4pcib6},
  note         = {Accessed: 2025-01-31}
}

@misc{hyperlane,
  author       = {Hyperlane},
  title        = {Hyperlane Official Website},
  howpublished = {\url{https://hyperlane.xyz/}},
  year         = {2025},
  note         = {Accessed: 2025-01-31}
}

@techreport{0x_white_paper,
  author       = {0x Project},
  title        = {0x White Paper},
  institution  = {0x Project},
  year         = {2025},
  month        = {January},
  url          = {https://github.com/0xProject/whitepaper/blob/master/0x_white_paper.pdf},
  note         = {Accessed: 2025-01-31}
}

@misc{0x_org,
  author       = {0x},
  title        = {0x Official Website},
  howpublished = {\url{https://0x.org/}},
  year         = {2025},
  note         = {Accessed: 2025-01-31}
}

@misc{layerzero_network,
  author       = {LayerZero},
  title        = {LayerZero Official Website},
  howpublished = {\url{https://layerzero.network/}},
  year         = {2025},
  note         = {Accessed: 2025-01-31}
}

\end{document}